\documentclass[preprint,aps,prd,showpacs,nofootinbib]{revtex4}

\usepackage{latexsym}
\usepackage{amsfonts}
\usepackage{amsbsy}
\usepackage{mathrsfs}

\begin{document}

\title{Hamilton-Jacobi theory for Hamiltonian systems with non-canonical
symplectic structures}
\date{\today}

\author{Aldo A. Mart\'{\i}nez-Merino}
\email{amerino@fis.cinvestav.mx} \affiliation{Departamento de F\'{\i}sica,
Cinvestav, Av. Instituto Polit\'ecnico Nacional 2508,\\ San Pedro Zacatenco,
07360, Gustavo A. Madero, Ciudad de M\'exico, M\'exico.}

\author{Merced Montesinos\footnote{Associate Member of the Abdus Salam
International Centre for Theoretical Physics, Trieste, Italy.}}
\email{merced@fis.cinvestav.mx} \affiliation{Departamento de F\'{\i}sica,
Cinvestav, Av. Instituto Polit\'ecnico Nacional 2508,\\ San Pedro Zacatenco,
07360, Gustavo A. Madero, Ciudad de M\'exico, M\'exico.}

\begin{abstract}
A proposal for the Hamilton-Jacobi theory in the context of the covariant
formulation of Hamiltonian systems is done. The current approach consists in
applying Dirac's method to the corresponding action which implies the
inclusion of second-class constraints in the formalism which are handled using
the procedure of Rothe and Scholtz recently reported. The current method is
applied to the nonrelativistic two-dimensional isotropic harmonic oscillator
employing the various symplectic structures for this dynamical system recently
reported.
\end{abstract}


\maketitle

\section{Introduction}
To set down the issue analyzed in this paper, we begin first with a brief
discussion of the standard treatment of Hamiltonian systems and after that
with a brief summary of what we call a genuine covariant description of
Hamiltonian dynamics.

\subsection{Canonical formulation of Hamiltonian systems}
In the standard treatment of Hamiltonian dynamics, the equations of motion are
written in the form \cite{Berndt}
\begin{eqnarray} \label{uhe}
{\dot q}^i = \frac{\partial H}{\partial p_i} \, , \quad {\dot p}_i = -
\frac{\partial H}{\partial q^i} \, , \quad i =1,2, \ldots , n \, ,
\end{eqnarray}
where $H$ is the Hamiltonian of the system, the variables $(q^i, p_i)$ are
canonically conjugate to each other in the sense that
\begin{eqnarray}\label{ccr}
\{ q^i , q^j \} =0 \, , \quad \{ q^i , p_j \} = \delta^i_j \, , \quad \{ p_i ,
p_j \} =0 \, ,
\end{eqnarray}
where $\{ , \}$ is the Poisson bracket defined by (summation convention is
used)
\begin{eqnarray}\label{upb}
\{ f , g \} = \frac{\partial f}{\partial q^i} \frac{\partial g}{\partial p_i}
- \frac{\partial f}{\partial p_i} \frac{\partial g}{\partial q^i} \, .
\end{eqnarray}

\subsection{Covariant formulation of Hamiltonian systems}
The symplectic geometry involved in the Hamiltonian description of mechanics
can clearly be appreciated if Eq. (\ref{uhe}) are written in the form
\begin{eqnarray}\label{che}
{\dot x}^{\mu} = \omega^{\mu\nu} \frac{\partial H}{\partial x^{\nu}} \, ,
\end{eqnarray}
with $ ( x^{\mu} ) = ( q^1, \ldots , q^n ; p_1 , \ldots , p_n )$ and
\begin{eqnarray}\label{usm}
\left ( \omega^{\mu\nu} \right ) & = & \left (
\begin{array}{cc}
0 & I \\
- I & 0
\end{array}
\right ) \, ,
\end{eqnarray}
where $0$ is the zero $n \times n$ matrix and $I$ is the unity $n\times n$
matrix. Moreover, Eq. (\ref{upb}) acquires the form
\begin{eqnarray}
\{ f , g \}= \frac{\partial f}{\partial x^{\mu}} \omega^{\mu\nu}
\frac{\partial g}{\partial x^{\nu}} \, ,
\end{eqnarray}
from which Eq. (\ref{ccr}) can be rewritten as
\begin{eqnarray}
\{ x^{\mu} , x^{\nu} \} = \omega^{\mu\nu} \, .
\end{eqnarray}
From this viewpoint, the coordinates $(x^{\mu})$ locally label the points $x$
of the phase space $\Gamma$ associated to the dynamical system on which the
symplectic structure $\omega=\frac12 \omega_{\mu\nu}(x) d x^{\mu} \wedge d
x^{\nu}$ is defined. The two-form $\omega$ is closed, i.e.,  $d\omega=0$ which
is equivalent to the fact that the Poisson bracket satisfies the Jacobi
identity \cite{Berndt}. Also $\omega$ is non-degenerate, i.e.,
$\omega_{\mu\nu} v^{\nu}=0$ implies $v^{\mu}=0$ which means that there exists
the inverse matrix $(\omega^{\mu\nu})$. The equations of motion of Eq.
(\ref{che}) are {\it covariant}\/ in the sense that they maintain their form
if the canonical coordinates are replaced by a completely arbitrary set of
coordinates in terms of which $(\omega^{\mu\nu})$ need not be given by Eq.
(\ref{usm}). Similarly, it is possible to retain the original coordinates
$(q^i,p_i)$ and still write the original equations of motion (\ref{uhe}) in
the Hamiltonian form (\ref{che}), but now employing alternative symplectic
structures $\omega^{\mu\nu} (x)$, distinct to that given in Eq. (\ref{usm}),
and taking as Hamiltonian any real function on $\Gamma$ which is a constant of
motion for the system. This means that the writing of the equations of motion
of a dynamical system in Hamiltonian form is {\it not} unique
\cite{Ger0,Ger1,Ger3,MonPRA03,MonGer04,mer}. More precisely, from the
covariant viewpoint of Hamiltonian dynamics, what is relevant in the
Hamiltonian formalism is the fact that the phase space $\Gamma$ is endowed
with a symplectic structure $\omega$ and a Hamiltonian $H$, not the fact of
singling out coordinates $q$'s and momenta $p$'s to label the points of
$\Gamma$ (see Refs. \cite{Ger0,Ger1,Ger3,MonPRA03,MonGer04,mer} where various
symplectic structures with respect to the same set of coordinates and for the
same dynamical system are discussed).

Before going on, it is convenient to remind the reader that the term {\it
covariant}\/ is also used in the context of the {\it covariant canonical
formalism}\/ of Ref. \cite{Witten} to refer precisely to the fact that what is
relevant in the Hamiltonian description of dynamics is the fact of having a
symplectic two-form on the phase space $\Gamma$ and not the fact of picking
out coordinates $q$'s and momenta $p$'s on it, as already mentioned. Even
though this observation is correct, the authors of Ref. \cite{Witten} obtain
the symplectic structure for a particular dynamical system using only its
equations of motion. This fact has generated the belief that the equations of
motion uniquely determine the symplectic geometry associated with any
Hamiltonian system, which is not true. In particular, the {\it space of
solutions}\/ can be endowed with more than one symplectic structure and no one
is more natural than the others (see Refs. \cite{MonPRA03,MonGer04,MonMon}).

On the other hand, in the same sense that the action
\begin{eqnarray}\label{uaction}
S [q^i , p_i ] = \int^{t_2}_{t_1} dt \left [ {\dot q}^i p_i - H(q,p,t) \right
] \, ,
\end{eqnarray}
provides the usual equations of motion (\ref{uhe}), the covariant form of
Hamilton's equations (\ref{che}) can be obtained from the action \cite{mer}
\begin{eqnarray}\label{covaction}
S [ x^{\mu} ] = \int^{t_2}_{t_1} d t \left [ \theta_{\mu} (x) \frac{d
x^{\mu}}{dt} - H(x,t) \right ] \, ,
\end{eqnarray}
provided ${\tilde \delta} S=0$ and ${\tilde \delta} x^{\mu} (t_1)=0=
{\tilde\delta} x^{\mu} (t_2)$ under the arbitrary configurational (or form)
variation of the variables $x^{\mu}$ at $t$ fixed, ${\tilde\delta} x^{\mu}$.
In fact,
\begin{eqnarray}
{\tilde\delta } S & = & \int^{t_2}_{t_1} dt \left ( \omega_{\nu\mu} (x) {\dot
x}^{\mu} - \frac{\partial H}{\partial x^{\nu}} \right ) {\tilde \delta}
x^{\nu} + \left ( \theta_{\mu} (x) {\tilde\delta} x^{\mu} \right )
\mid^{t_2}_{t_1} \, ,
\end{eqnarray}
where $\omega$ is the symplectic two-form
\begin{eqnarray}\label{jaja}
\omega & = & \frac12 \omega_{\mu\nu} (x) d x^{\mu} \wedge d x^{\nu}\, , \quad
\omega_{\mu\nu}  = \partial_{\mu} \theta_{\nu} - \partial_{\nu} \theta_{\mu}\,
.
\end{eqnarray}
Equivalently, $\omega= d \theta$ where $\theta= \theta_{\mu} d x^{\mu}$ is the
symplectic potential. It is convenient to make clear some aspects involved
with the boundary conditions ${\tilde\delta} x^{\mu} (t_1)=0 = {\tilde\delta}
x^{\mu} (t_2)$ employed in Hamilton's principle. Due to the fact that there
are $2n$ coordinates $x^{\mu}$, one must fix only $2n$ conditions at the time
boundary (which might be $n$ conditions at $t=t_1$ and $n$ conditions at
$t=t_2$), otherwise the system might be over-determined, in which case the
system might not evolve from $t_1$ to $t_2$. For instance, in the case when
$\theta = p_i d q^i$, it is clear that one can arbitrarily choose
${\tilde\delta} q^i (t_1)=0$ and ${\tilde\delta} q^i (t_2)=0$. However, in the
generic case, namely, when $\theta= \theta_{\mu} (x) d x^{\mu}$ one can still
arbitrarily choose ${\tilde\delta} x^{\mu} (t_1)=0$ at $t=t_1$. Nevertheless,
even though ${\tilde\delta} x^{\mu} (t_2)=0 $ still holds, $x^{\mu} (t_2)$
cannot be arbitrarily chosen, but it is fixed by the conditions on $x^{\mu}$
at $t_1$ in order for the system to evolve from $t_1$ to $t_2$.

Up to here the description of both the canonical and the covariant description
of Hamiltonian systems. Now, in the framework of the canonical description of
Hamiltonian systems the Hamilton-Jacobi theory of these type of systems is
built following the usual procedure (see, for instance, Ref. \cite{Berndt}).
Nevertheless, suppose that one wants to remain in the framework of the
covariant description of Hamiltonian dynamics. The question is, is there a
well-defined Hamilton-Jacobi theory if symplectic structures alternative to
the usual ones are used? The answer is in the affirmative and the
corresponding formalism is developed in next section (see also Ref.
\cite{merino} for more details).

\section{Hamilton-Jacobi theory}\label{hjt}
The starting point is the action of Eq. (\ref{covaction}). Using Dirac's
method \cite{Dirac}, all the variables $x^{\mu}$ of which the action
(\ref{covaction}) depends functionally on are taken as configuration
variables. The next step is to define the momenta $\pi_{\mu}$ canonically
conjugate to the $x^{\mu}$. By definition
\begin{eqnarray}
\pi_{\mu}:= \theta_{\mu} (x) \, , \quad \mu=1,2,\ldots, 2n \, ,
\end{eqnarray}
which lead to the primary constraints
\begin{eqnarray}\label{primaryconst}
\chi_{\mu} := \pi_{\mu} - \theta_{\mu} (x) = 0 \, , \quad \mu=1,2,\ldots , 2n
\, .
\end{eqnarray}
In this way, the points of the extended phase space $\Gamma_{ext}$ are
labelled with the coordinates $(x^{\mu}, \pi_{\nu})$ and the symplectic
structure in these coordinates is given by
\begin{eqnarray}\label{extse}
\{ x^{\mu}, x^{\nu} \} = 0\, , \quad \{ x^{\mu} , \pi_{\nu} \} =
\delta^{\mu}_{\nu}\, , \quad \{ \pi_{\mu} , \pi_{\nu} \} = 0 \, ,
\end{eqnarray}
or, equivalently $\Omega=d \pi_{\mu} \wedge d x^{\mu}$. Performing the
Legendre transformation, the canonical Hamiltonian $H_c$ is simply $H$, $H_c=
\pi_{\mu}{\dot x}^{\mu} - L = \pi_{\mu} {\dot x}^{\mu} - \left ( \theta_{\mu}
{\dot x}^{\mu} -H \right ) = H$. Therefore, following Dirac's procedure, the
action principle is promoted to
\begin{eqnarray}\label{firstext}
S [ x^{\mu}, \pi_{\mu}, \Lambda^{\mu} ] = \int^{t_2}_{t_1} dt \left [ {\dot
x}^{\mu} \pi_{\mu} - H(x,t) - \Lambda^{\mu} \chi_{\mu} \right ]\, ,
\end{eqnarray}
where $\Lambda^{\mu}$ are Lagrange multipliers. From the action of Eq.
(\ref{firstext}), the dynamical equations
\begin{eqnarray}\label{dyneqs}
{\dot x}^{\mu} & = & \Lambda^{\mu} \, , \nonumber\\
{\dot \pi}_{\mu} & = & - \frac{\partial H}{\partial x^{\mu}} + \Lambda^{\nu}
\frac{\partial \theta_{\nu}}{\partial x^{\mu}} \, ,
\end{eqnarray}
together with the constraint (\ref{primaryconst}) are obtained.

By using Eq. (\ref{dyneqs}), the evolution of the constraints $\chi_{\mu}$ is
computed
\begin{eqnarray}
{\dot \chi}_{\mu} & = & {\dot \pi}_{\mu}- {\dot \theta}_{\mu}  \nonumber\\
& = & {\dot \pi}_{\mu} - \frac{\partial \theta_{\mu}}{\partial x^{\nu}} {\dot
x}^{\nu} \nonumber\\
& = & - \frac{\partial H}{\partial x^{\mu}} + \omega_{\mu\nu} \Lambda^{\nu} \,
,
\end{eqnarray}
where $\omega_{\mu\nu}(x)$ in last equality is the same one given in Eq.
(\ref{jaja}). Therefore, ${\dot \chi}_{\mu} \approx 0$ fixes the Lagrange
multipliers
\begin{eqnarray}\label{Lag}
\Lambda^{\nu} \approx \omega^{\nu\mu} \frac{\partial H}{\partial x^{\mu}} \, ,
\end{eqnarray}
and no more constraints arise. Moreover, the constraints $\chi_{\mu}$ are {\it
second class}. In fact, using the symplectic structure on $\Gamma_{ext}$ given
in Eq. (\ref{extse}) one has
\begin{eqnarray}
\{ \chi_{\mu} , \chi_{\nu} \} = \omega_{\mu\nu} (x) \, ,
\end{eqnarray}
which, by hypothesis, has non-vanishing determinant (see Eq. (\ref{jaja})).
Furthermore, by inserting the expressions of the Lagrange multipliers
(\ref{Lag}) into the Hamiltonian $H_c + \Lambda^{\mu} \chi_{\mu}$ one gets the
{\it first-class} Hamiltonian $H'= H + \chi_{\mu} \omega^{\mu\nu}
\frac{\partial H}{\partial x^{\nu}}$. In fact, one readily verifies that
\begin{eqnarray}
\{ H' , \chi_{\mu} \} = \chi_{\alpha} \{ \omega^{\alpha\beta} \frac{\partial
H}{\partial x^{\beta}} , \chi_{\mu} \} \approx 0 \, ,
\end{eqnarray}
so the usual structure between the first-class Hamiltonian and second-class
constraints is satisfied \cite{henneaux}.

Finally, the action principle acquires the form
\begin{eqnarray}\label{finala}
S[x^{\mu},\pi_{\mu}, \lambda^{\mu} ] = \int^{t_2}_{t_1} dt \left [ {\dot
x}^{\mu} \pi_{\mu} - H' - \lambda^{\mu} \chi_{\mu} \right ]\, ,
\end{eqnarray}
where $\lambda^{\mu}$ are new Lagrange multipliers.

In summary, the application of Dirac's method to the action (\ref{covaction})
which is associated with an unconstrained Hamiltonian system and described
with non-canonical symplectic structures implies the introduction of
second-class constraints in the extended phase space $\Gamma_{ext}$ which is
endowed with a canonical symplectic structure. Therefore, the original problem
of building the Hamilton-Jacobi theory for unconstrained Hamiltonian systems
described by non-canonical symplectic structures has been transformed into the
one of building the Hamilton-Jacobi theory for systems with second-class
constraints with respect to a canonical symplectic structure. Fortunately,
there is a proposal to build the Hamilton-Jacobi theory when second class are
involved \cite{rothe}. In such a paper, the analysis is restricted to
second-class constraints linear in the coordinates and in the momenta.
However, there is no need to restrict the analysis to this particular kind of
second-class constraints if the analysis is locally carried out (see pages 46
and 64 of Ref. \cite{henneaux}). The procedure of Ref. \cite{rothe} consists,
essentially, in making a $t$-independent canonical transformation from the
original canonical variables $(x^{\mu}, \pi_{\nu})$ which label the point of
$\Gamma_{ext}$ to new canonical variables $(X^{\mu},\Pi_{\nu})$ in terms of
which the original second-class constraints $\chi_{\mu}$ become canonically
conjugate pairs $(Q^a,P_b)$ which form part of the new set of canonical
variables $(X^{\mu};\Pi_{\nu}) \equiv (q^r_{\ast}, Q^a ; p^{\ast}_s, P_b)$.
Once this is done, the original first-class Hamiltonian $H'$ is rewritten in
terms of the new canonical variables
\begin{eqnarray}
{\tilde H} \left (  q_{\ast}, Q, p^{\ast}, P \right ) :=  H' \left ( x \left (
q_{\ast}, Q, p^{\ast}, P \right ) , \pi \left ( q_{\ast}, Q, p^{\ast}, P
\right ), t \right )
\end{eqnarray}
By setting the second-class constraints strongly equal to zero, $Q^a=0$ and
$P_b=0$, in the Hamiltonian ${\tilde H}$, ${\tilde H}(q_{\ast}, Q=0, p^{\ast},
P=0 ) =: {\hat H}(q_{\ast}, p_{\ast},t)$, the corresponding Hamilton-Jacobi
equation arises
\begin{eqnarray}\label{hjequation}
\frac{\partial S}{\partial t} + {\hat H}  \left ( q^{\ast}, \frac{\partial
S}{\partial q_{\ast} }, t  \right ) = 0 \, .
\end{eqnarray}

\section{Examples}
Now, the implementation of the procedure developed in the Section \ref{hjt} is
carried out. The starting point is the equations of motion
\begin{eqnarray}\label{ho}
{\dot x} = \frac{p_x}{m}\, , \quad {\dot y}= \frac{p_y}{m}\, , \quad {\dot
p}_x = - m \varpi^2 x \, , \quad {\dot p}_y = - m \varpi^2 y \, ,
\end{eqnarray}
for the two-dimensional isotropic non-relativistic harmonic oscillator. Here,
the dot ``$\cdot$" stands for the total derivative with respect to the
Newtonian time $t$, $m$ is the mass of the particle and $\varpi$ the angular
frequency. The canonical formulation of the equations of motion (\ref{ho})
consists in taking $(x^{\mu})=(x^1,x^2,x^3,x^4)=(x,y,p_x,p_y)$ as coordinates
to label the points $x$ of the phase space $\Gamma=\mathbb{R}^4$ of the system
together with the symplectic structure (\ref{usm}) and Hamiltonian
$H=\frac{1}{2m} \left ( (p_x)^2 + m^2 \varpi^2 x^2 + (p_y)^2 + m^2 \varpi^2
y^2\right )$.

\subsection{first case}
Alternatively, according to the covariant formulation of Hamiltonian systems,
the equations of motion (\ref{ho}) can be written in Hamiltonian form by
taking $(x^{\mu})=(x^1,x^2,x^3,x^4)=(x,y,p_x,p_y)$ as coordinates to label the
points $x$ of the phase space $\Gamma=\mathbb{R}^4$ of the system together
with the symplectic structure and Hamiltonian \cite{Ger3}
\begin{eqnarray}\label{case1}
\left ( \omega^{\mu\nu} \right ) = \left (
\begin{array}{cccc}
0 & 0 & 0 & 1 \\
0 & 0 & 1 & 0 \\
0 & -1 & 0 & 0 \\
-1 & 0 & 0 & 0
\end{array}
\right )\, , \quad H = \frac{p_x p_y}{m} + m \varpi^2 x y \, .
\end{eqnarray}
In fact, one easily verifies that by inserting the symplectic structure and
Hamiltonian given in Eq. (\ref{case1}) into Eq. (\ref{che}), the equations of
motion (\ref{ho}) are obtained. Moreover, the symplectic structure of Eq.
(\ref{case1}) can be obtained from the potential 1-form $\theta$, $\omega= d
\theta= d \left ( p_y d x + p_x d y \right ) = d p_y \wedge d x + d p_x \wedge
d y $, so it is possible to give the action principle \cite{MonMon}
\begin{eqnarray}
S [x,y,p_x,p_y ] = \int^{t_2}_{t_1} dt  \left [ p_y {\dot x} + p_x {\dot y} -
\left ( \frac{p_x p_y}{m} + m \varpi^2 x y \right ) \right ] \, ,
\end{eqnarray}
which provides this Hamiltonian formulation.

Following the procedure described in the Section \ref{hjt}, all the variables
$(x^{\mu})=(x,y,p_x,p_y)$ are taken as configuration variables. Then, Dirac's
method calls for the definition of the momenta $(\pi_{\mu})=(\pi_1, \pi_2 ,
\pi_3 , \pi_4)$ canonically conjugate to $(x^{\mu})$; respectively. So, the
points of the extended phase space $\Gamma_{ext}=\mathbb{R}^8$ are labelled
with $(x^{\mu}, \pi_{\nu} )$. From the definition of the momenta, one has
\begin{eqnarray}\label{scc1}
\chi_1 & := & \pi_1 - p_y = 0 \, , \nonumber\\
\chi_2 & := & \pi_2 - p_x = 0 \, , \nonumber\\
\chi_3 & := & \pi_3 =0 \, , \nonumber\\
\chi_4 & := & \pi_4 =0 \, ,
\end{eqnarray}
which are second-class constraints. The systematic implementation of the
procedure leads to the action principle (\ref{finala}) with first-class
Hamiltonian $H'$
\begin{eqnarray}\label{ham1}
H' = m \varpi^2 x y - \frac{p_x p_y}{m} + \frac{1}{m} \left ( p_x \pi_1 + p_y
\pi_2 \right ) -  m \varpi^2 \left (  x \pi_3 + y \pi_4  \right )\, .
\end{eqnarray}
In fact,
\begin{eqnarray}
\{ \chi_1 , H' \} = m \varpi^2 \chi_3 \, , \nonumber\\
\{ \chi_2 , H' \} = m \varpi^2 \chi_4 \, , \nonumber\\
\{ \chi_3 , H' \} = - \frac{1}{m} \chi_1 \, , \nonumber\\
\{ \chi_4 , H' \} = - \frac{1}{m} \chi_2 \, .
\end{eqnarray}
Next, a canonical transformation in $\Gamma_{ext}={\mathbb{R}}^8$ from
$(x,y,p_x,p_y;\pi_1,\pi_2,\pi_3,\pi_4)$ to $(q^1_{\ast}, q^2_{\ast}, Q^1, Q^2;
p^{\ast}_1, p^{\ast}_2, P_1, P_2)$ such that the original second-class
constraints (\ref{scc1}) form canonical pairs of the new set of canonical
variables is performed. The new canonical pairs are $(q^1_{\ast},p^{\ast}_1)$,
$(q^2_{\ast},p^{\ast}_2)$, $(Q^1,P_1)$, and $(Q^2,P_2)$. The relationship
between these canonical variables and the original ones is
\begin{eqnarray}
q^1_{\ast} & = & x - \pi_4 \, , \quad p^{\ast}_1 = \pi_1 \, , \quad \,\,
q^2_{\ast} = y - \pi_3 \, , \quad p^{\ast}_4 = \pi_2 \, , \nonumber\\
Q^1 & = & \chi_3 \, , \quad \quad \,\,\,\, P_1 =\chi_2 \, , \quad \, Q^2 =
\chi_4 \, , \quad \quad \,\,\,\, P_2 = \chi_1 \, .
\end{eqnarray}
Hence, the inverse transformation is given by
\begin{eqnarray}\label{inv1}
x & = & q^1_{\ast} + Q^1 \, , \quad \pi_1 = p^{\ast}_1 \, , \quad \,\,\,\, y =
q^2_{\ast} + Q^2 \, , \quad \pi_2 = p^{\ast}_2 \, , \nonumber\\
p_x & = & p^{\ast}_2 - P_2 \, , \quad \, \pi_3 = Q^2 \, , \quad p_y =
p^{\ast}_1 - P_1 \, , \quad \pi_4 = Q^1 \, .
\end{eqnarray}
Then, in terms of the new set of canonical variables the Hamiltonian
(\ref{ham1}) acquires the form
\begin{eqnarray}
{\tilde H} & = & H' \left ( x^{\mu} (q_{\ast},p^{\ast},Q,P), \pi_{\mu}
(q_{\ast},p^{\ast},Q,P) \right ) \nonumber\\
& = & \frac{1}{m} \left [ \left ( p^{\ast}_2 - P_2 \right ) p^{\ast}_1 + \left
( p^{\ast}_1 - P_1 \right ) p^{\ast}_2 - \left ( p^{\ast}_2 - P_2 \right )
\left ( p^{\ast}_1 - p_1 \right ) \right ] \nonumber\\
& & \mbox{} + m \varpi^2 \left [ \left ( q^1_{\ast} + Q^1 \right ) \left (
q^2_{\ast} + Q^2 \right ) - \left ( q^1_{\ast} + Q^1 \right ) Q^2 - \left (
q^2_{\ast} + Q^2 \right ) Q^1 \right ] \, .
\end{eqnarray}
By applying the procedure of Ref. \cite{rothe}, which means to set $Q^a=0,
P_a=0$, $a=1,2$, together with $p^{\ast}_r = \frac{\partial S}{\partial
q^r_{\ast}}$, $r=1,2$, the Hamilton-Jacobi equation is obtained
\begin{eqnarray}
\frac{\partial S}{\partial t} + \frac{1}{m} \frac{\partial S}{\partial
q^1_{\ast}} \frac{\partial S}{\partial q^2_{\ast}} + m \varpi^2 q^1_{\ast}
q^2_{\ast} & = & 0 \, .
\end{eqnarray}
A complete solution of last equation is given by
\begin{eqnarray}
S & = & \frac{m\varpi}{\sin{\varpi t}} \left ( \left ( q^1_{\ast} q^2_{\ast} +
q^1_{\ast 0} q^2_{\ast 0} \right ) \cos{\varpi t }  - \left ( q^1_{\ast}
q^2_{\ast 0} + q^1_{\ast 0} q^2_{\ast} \right ) \right ) \, ,
\end{eqnarray}
where $q^1_{\ast 0}$ and $q^2_{\ast 0}$ are integration constants. Therefore,
the momenta $p^{\ast}_r$ and $p^{\ast}_{r0}$ canonically conjugate to
$q^r_{\ast}$ and $q^r_{\ast 0}$ are obtained from $p^{\ast}_r = \frac{\partial
S }{\partial q^r_{\ast}}$ and $- p^{\ast}_{r0} = \frac{\partial S }{\partial
q^r_{\ast 0}}$; respectively
\begin{eqnarray}
p^{\ast}_1 & = & \frac{m\varpi}{\sin{\varpi t}} \left ( q^2_{\ast} \cos{\varpi
t } - q^2_{\ast 0} \right )\, , \nonumber\\
p^{\ast}_2 & = & \frac{m \varpi}{\sin{\varpi t}} \left ( q^1_{\ast}
\cos{\varpi t} - q^1_{\ast 0} \right ) \, , \nonumber\\
- p^{\ast}_{10} & = & \frac{m \varpi}{\sin{\varpi t} } \left ( q^2_{\ast 0 }
\cos{\varpi t} - q^2_{\ast} \right ) \, , \nonumber\\
- p^{\ast}_{20} & = & \frac{m\varpi}{\sin{\varpi t}} \left ( q^1_{\ast 0}
\cos{\varpi t} - q^1_{\ast} \right ) \, .
\end{eqnarray}
By plugging these equations together with the constraints $Q^1=0$, $Q^2=0$,
$P_1=0$, and $P_2=0$ into (\ref{inv1}) the solution to the original equations
of motion is obtained
\begin{eqnarray}\label{solution}
x & = & x_0 \cos{\varpi t} + \frac{p_{x0}}{m\varpi} \sin{\varpi t} \, , \quad
p_x = - m\varpi x_0 \sin{\varpi t} + p_{x0} \cos{\varpi t} \, , \nonumber\\
y & = & y_0 \cos{\varpi t} + \frac{p_{y0}}{m\varpi} \sin{\varpi t} \, , \quad
\, p_y = - m \varpi y_0 \sin{\varpi t} + p_{y0} \cos{\varpi t} \, .
\end{eqnarray}

\subsection{Second case}
Similarly, according to the covariant viewpoint of Hamiltonian dynamics, the
equations of motion (\ref{ho}) can be written in Hamiltonian form by taking as
symplectic structure and Hamiltonian \cite{Ger3}
\begin{eqnarray}\label{case2}
\left ( \omega^{\mu\nu} \right ) = \left (
\begin{array}{cccc}
0 & 0 & -1 & 0 \\
0 & 0 & 0 & 1 \\
1 & 0 & 0 & 0 \\
0 & -1 & 0 & 0
\end{array}
\right )\, , H = \frac{1}{2m} \left ( (p_y)^2 - (p_x)^2 \right ) + \frac{m
\varpi^2}{2} \left ( y^2 - x^2 \right ) \, ,
\end{eqnarray}
keeping the same coordinates $(x^{\mu})=(x^1,x^2,x^3,x^4)=(x,y,p_x,p_y)$ to
label the points $x$ of the phase space $\Gamma={\mathbb{R}}^4$. The
symplectic structure of Eq. (\ref{case2}) can be obtained from the potential
1-form $\theta$, $\omega=d\theta= d \left ( x d p_x +p_y d y \right )= - d p_x
\wedge dx + d p_y \wedge dy$, so it is possible to give the action principle
\begin{eqnarray}
S[x,y,p_x,p_y] & = & \int^{t_2}_{t_1} dt \left [ x {\dot p}_x + p_y {\dot y}
\right. \nonumber\\
& & \mbox{} - \left. \left ( \frac{1}{2m} \left ( (p_y)^2 - (p_x)^2 \right ) +
\frac{m \varpi^2}{2} \left ( y^2 - x^2 \right ) \right )  \right ]\, .
\end{eqnarray}
The definition of the momenta $(\pi_{\mu})=(\pi_1,\pi_2,\pi_2,\pi_4)$
canonically conjugate to $(x^{\mu})=(x,y,p_x,p_y)$ implies the inclusion of
second-class constraints
\begin{eqnarray}\label{scc2}
\chi_1 & := & \pi_1 =0 \, , \nonumber\\
\chi_2 & := & \pi_2 - p_y =0 \, , \nonumber\\
\chi_3 & := & \pi_3 - x =0 \, , \nonumber\\
\chi_4 & := & \pi_4 =0 \, ,
\end{eqnarray}
and the first-class Hamiltonian
\begin{eqnarray}\label{fch2}
H' & = & \frac{m \varpi^2}{2} \left ( x^2 + y^2 \right ) - \frac{1}{2m} \left
( (p_x)^2 + (p_y)^2 \right ) \nonumber\\
& & \mbox{} + \frac{1}{m} \left ( p_x \pi_1 + p_y \pi_2 \right ) - m \varpi^2
\left ( x \pi_3 + y \pi_4 \right )\, .
\end{eqnarray}
The Poisson brackets between the second-class constraints and $H'$ are
\begin{eqnarray}
\{ \chi_1 , H' \} = m \varpi^2 \chi_3 \, , \nonumber\\
\{ \chi_2 , H' \} = m \varpi^2 \chi_4 \, , \nonumber\\
\{ \chi_3 , H' \} = - \frac{1}{m} \chi_1 \, , \nonumber\\
\{ \chi_4 , H' \} = - \frac{1}{m} \chi_2 \, ,
\end{eqnarray}
which have the usual structure \cite{henneaux}. So, the coordinates
$(x^{\mu},\pi_{\nu})$ label the points of the extended phase space
$\Gamma_{ext}={\mathbb{R}}^8$ which is endowed with the symplectic structure
$\Omega=d\pi_{\mu} \wedge d x^{\mu}$.

Next, a canonical transformation in $\Gamma_{ext}={\mathbb{R}}^8$ from
$(x,y,p_x,p_y;\pi_1,\pi_2,\pi_3,\pi_4)$ to $(q^1_{\ast}, q^2_{\ast}, Q^1, Q^2;
p^{\ast}_1, p^{\ast}_2, P_1, P_2)$ such that the original second-class
constraints (\ref{scc2}) form canonical pairs of the new set of canonical
variables is performed. The new canonical pairs are $(q^1_{\ast},p^{\ast}_1)$,
$(q^2_{\ast},p^{\ast}_2)$, $(Q^1,P_1)$, and $(Q^2,P_2)$. The relationship
between these canonical variables and the original ones is
\begin{eqnarray}
q^1_{\ast} & = & p_x - \pi_1 \, , \quad p^{\ast}_1 = \pi_3 \, , \quad \,\,
q^2_{\ast} = y - \pi_4 \, ,\quad p^{\ast}_2 = \pi_2 \, , \nonumber\\
Q^1 & = & \chi_1 \, , \quad \quad \quad P_1 = \chi_3 \, , \quad Q^2 = \chi_4
\, , \quad \quad \,\,\, P_2 = \chi_2 \, ,
\end{eqnarray}
with the corresponding inverse transformation
\begin{eqnarray}\label{inv2}
x & = & p^{\ast}_1 - P_1 \, ,\quad \pi_1 = Q^1 \, , \quad \, \, y = q^2_{\ast}
+ Q^2
\, , \quad \pi_2 = p^{\ast}_2 \, ,\nonumber\\
p_x & = & q^1_{\ast} + Q^1 \, , \quad \pi_3 = \pi^{\ast}_1 \, , \quad p_y =
p^{\ast}_2 - P_2 \, , \quad \, \pi_4 = Q^2 \, .
\end{eqnarray}
Inserting these into the first-class Hamiltonian (\ref{fch2})
\begin{eqnarray}
{\tilde H} & = & \frac{m \varpi^2}{2} \left[ \left ( q^2_{\ast} + Q^2 \right
)^2 + \left ( p^{\ast}_1 - P_1 \right )^2 \right ]  - \frac{1}{2m} \left [
\left ( p^{\ast}_2 - P_2 \right )^2 + \left
( q^1_{\ast} + Q^1 \right )^2 \right ] \nonumber\\
& & \mbox{} + \frac{1}{m} \left [ \left ( q^1_{\ast} + Q^1 \right ) Q^1 +
\left ( p^{\ast}_2 - P_2 \right ) p^{\ast}_2 \right ] \nonumber\\
& & \mbox{} - m \varpi^2 \left [ \left ( p^{\ast}_1 - P_1 \right ) p^{\ast}_1
+ \left ( q^2_{\ast} + Q^2 \right ) Q^2 \right ] \, .
\end{eqnarray}
By inserting $Q^a=0$, $P_a=0$, $a=1,2$ and $p^{\ast}_r = \frac{\partial
S}{\partial q^r_{\ast}}$, $r=1,2$ into last expression, the Hamilton-Jacobi
equation arises
\begin{eqnarray}
\frac{\partial S}{\partial t} + \frac{1}{2m} \left [ \left ( \frac{\partial S
}{\partial q^2_{\ast}} \right )^2 - (q^1_{\ast})^2  \right ] - \frac{m
\varpi^2}{2} \left [ \left ( \frac{\partial S }{\partial q^1_{\ast}} \right
)^2 - (q^2_{\ast})^2 \right ] = 0 \, .
\end{eqnarray}
A complete solution of this equation is given by
\begin{eqnarray}
S & = & \frac{m \varpi}{2 \sin{\varpi t}} \left [ \left ( (q^2_{\ast})^2 +
(q^2_{\ast 0})^2 - \frac{(q^1_{\ast})^2}{m^2 \varpi^2} - \frac{(q^1_{\ast 0
})^2}{m \varpi^2} \right ) \cos{\varpi t} \right. \nonumber\\
& & \mbox{} \left. - 2 \left ( q^2_{\ast} q^2_{\ast 0} - \frac{q^1_{\ast}
q^1_{\ast 0} }{m^2 \varpi^2} \right ) \right ]\, ,
\end{eqnarray}
where $q^1_{\ast 0}$ and $q^2_{\ast 0}$ are integration constants. Therefore,
the momenta $p^{\ast}_r$ and $p^{\ast}_{r0}$ canonically conjugate to
$q^r_{\ast}$ and $q^r_{\ast 0}$ are obtained from $p^{\ast}_r = \frac{\partial
S }{\partial q^r_{\ast}}$ and $- p^{\ast}_{r0} = \frac{\partial S }{\partial
q^r_{\ast 0}}$; respectively
\begin{eqnarray}
p^{\ast}_1 & = & - \frac{1}{m\varpi \sin{\varpi t}} \left ( q^1_{\ast}
\cos{\varpi t} - q^1_{\ast 0} \right ) \, , \nonumber\\
p^{\ast}_2 & = & \frac{m \varpi}{\sin{\varpi t}} \left ( q^2_{\ast}
\cos{\varpi t} - q^2_{\ast 0} \right )\, , \nonumber\\
- p^{\ast}_{10} & = & - \frac{1}{m\varpi \sin{\varpi t}} \left ( q^1_{\ast 0}
\cos{\varpi t} - q^1_{\ast} \right ) \, , \nonumber\\
- p^{\ast}_{20} & = & \frac{m\varpi}{\sin{\varpi t}} \left ( q^2_{\ast 0}
\cos{\varpi t} - q^2_{\ast} \right ) \, .
\end{eqnarray}
By using these equations and inserting $Q^a=0$, $P_a=0$, $a=1,2$ into
(\ref{inv2}) the solution (\ref{solution}) is obtained.

\subsection{Third case}
The equations of motion (\ref{ho}) can be written in Hamiltonian form by
taking as symplectic structure and Hamiltonian \cite{Ger3}
\begin{eqnarray}\label{case3}
\left ( \omega^{\mu\nu} \right ) = \left (
\begin{array}{cccc}
0 & - \frac{1}{m\varpi} & 0 & 0 \\
\frac{1}{m\varpi} & 0 & 0 & 0 \\
0 & 0 & 0 & - m \varpi \\
0 & 0 & m \varpi & 0
\end{array}
\right )\, , \quad H = \varpi \left ( x p_y - y p_x \right ) \, ,
\end{eqnarray}
where $(x^{\mu})=(x^1,x^2,x^3,x^4)=(x,y,p_x,p_y)$ label the points of the
phase space $\Gamma={\mathbb{R}}^4$. The symplectic structure (\ref{case3})
can be obtained from the potential 1-form $\theta$, $\omega= d \theta = d
\left ( - m \varpi y dx + \frac{1}{m\varpi} p_x d p_y \right )= m \varpi dx
\wedge dy + \frac{1}{m \varpi} d p_x \wedge d p_y $, so it is possible to give
the action principle
\begin{eqnarray}
S [ x,y,p_x,p_y ] & = & \int^{t_2}_{t_1} dt \left [ - m \varpi y {\dot x} +
\frac{p_x}{m \varpi} {\dot p}_y - \varpi \left ( x p_y - y p_x \right ) \right
]\, .
\end{eqnarray}
From the definition of the momenta $(\pi_{\mu})=(\pi_1,\pi_2,\pi_3,\pi_4)$
canonically conjugate to $(x^{\mu})$ one has that $\Omega=d \pi_{\mu} \wedge d
x^{\mu}$ is the symplectic structure on the extended phase space
$\Gamma_{ext}={\mathbb{R}}^8$ and that the second-class constraints
\begin{eqnarray}\label{scc3}
\chi_1 & := & \frac{\pi_1}{m \varpi} + y =0\, , \nonumber\\
\chi_2 & := & \pi_2 =0\, , \nonumber\\
\chi_3 & := & \pi_3 =0\, , \nonumber\\
\chi_4 & := & m \varpi \pi_4 - p_x =0 \, ,
\end{eqnarray}
arise together with the first-class Hamiltonian $H'$
\begin{eqnarray}\label{fch3}
H' = \varpi (x p_y + y p_x ) + \frac{1}{m} \left ( p_x \pi_1 + p_y \pi_2
\right ) - m \varpi^2 \left ( x \pi_3 + y \pi_4 \right ) \, ,
\end{eqnarray}
which satisfy
\begin{eqnarray}
\{ \chi_1 , H' \} & = & m \varpi^2 \chi_3 \, , \nonumber\\
\{ \chi_2 , H' \} & = & m \varpi^2 \chi_4 \, , \nonumber\\
\{ \chi_3 , H' \} & = & - \frac{1}{m} \chi_1 \, , \nonumber\\
\{  \chi_4 , H' \} & = & - \frac{1}{m} \chi_2 \, ,
\end{eqnarray}
as expected \cite{henneaux}.

Next, a canonical transformation in $\Gamma_{ext}={\mathbb{R}}^8$ from
$(x,y,p_x,p_y;\pi_1,\pi_2,\pi_3,\pi_4)$ to $(q^1_{\ast}, q^2_{\ast}, Q^1, Q^2;
p^{\ast}_1, p^{\ast}_2, P_1, P_2)$ such that the original second-class
constraints (\ref{scc3}) form canonical pairs of the new set of canonical
variables is performed. The new canonical pairs are $(q^1_{\ast},p^{\ast}_1)$,
$(q^2_{\ast},p^{\ast}_2)$, $(Q^1,P_1)$, and $(Q^2,P_2)$. The relationship
between these canonical variables and the original ones is
\begin{eqnarray}
q^1_{\ast} & = & m \varpi x + \pi_2 \, , \quad p^{\ast}_1 =
\frac{\pi_1}{m\varpi} \, , \quad q^2_{\ast} = \frac{p_y}{m\varpi} - \pi_3 \, ,
\quad p^{\ast}_2 = m \varpi \pi_4 \, , \nonumber\\
Q^1 & = & \chi_1 \, , \quad \quad\quad\quad \,\, P_1 = \chi_2 \, , \quad
\,\,\, Q^2 = \chi_3 \, , \quad \quad \quad \quad P_2 = \chi_4 \, ,
\end{eqnarray}
with inverse transformation
\begin{eqnarray}\label{inv3}
x & = & \frac{1}{m \varpi} \left ( q^1_{\ast} - P_1 \right ) \, ,  \pi_1 = m
\varpi p^{\ast}_1 \, , \quad y = Q^1 - p^{\ast}_1 \, , \quad \quad\quad \pi_2
= P_1
\, , \nonumber\\
p_x & = & p^{\ast}_2 - P_2 \, , \quad \quad\quad \,\, \pi_3 = Q^2 \, , \quad
\quad p_y = m \varpi \left ( q^2_{\ast} + Q^2 \right ) \, ,\pi_4 =
\frac{p^{\ast}_2}{m\varpi} \, .
\end{eqnarray}
Then, in terms of the new set of canonical variables the first-class
Hamiltonian $H'$ (\ref{fch3}) acquires the form
\begin{eqnarray}
{\tilde H} & = & \varpi \left [ \left ( q^1_{\ast} - P_1 \right ) \left (
q^2_{\ast} + Q^2 \right ) + \left ( Q^1 - p^{\ast}_1 \right ) \left (
p^{\ast}_2 - p_2 \right ) + \left ( p^{\ast}_2 - P_2 \right ) p^{\ast}_1
\right. \nonumber\\
& & \mbox{} \left. + \left ( q^2_{\ast} + Q^2 \right ) P_1 - \left (
q^1_{\ast} - P_1 \right ) Q^2 - \left ( Q^1 - p^{\ast}_1 \right ) p^{\ast}_2
\right ]\, .
\end{eqnarray}
By applying the procedure of Ref. \cite{rothe}, which means to set $Q^a=0$,
$P_a=0$, $a=1,2$ and $p_r = \frac{\partial S}{\partial q^r_{\ast}}$, $r=1,2$,
the Hamilton-Jacobi equation arises
\begin{eqnarray}
\frac{\partial S}{\partial t} + \varpi \left ( \frac{\partial S}{\partial
q^1_{\ast}} \frac{\partial S}{\partial q^2_{\ast}} + q^1_{\ast} q^2_{\ast}
\right ) & = & 0 \, .
\end{eqnarray}
A complete solution of this equation is
\begin{eqnarray}
S = \frac{1}{\sin{\varpi t}} \left [ \left ( q^1_{\ast} q^2_{\ast} + q^1_{\ast
0} q^2_{\ast 0} \right ) \cos{\varpi t} - \left ( q^1_{\ast} q^2_{\ast 0} +
q^1_{\ast 0} q^2_{\ast} \right ) \right ]\, ,
\end{eqnarray}
where $q^1_{\ast 0}$ and $q^2_{\ast 0}$ are integration constants. Hence, the
momenta $p^{\ast}_r$ and $p^{\ast}_{r0}$ canonically conjugate to $q^r_{\ast}$
and $q^r_{\ast 0}$, are obtained from $p^{\ast}_r = \frac{\partial S}{\partial
q^r_{\ast}}$ and $- p^{\ast}_{r0}= \frac{\partial S}{\partial q^r_{\ast 0}}$;
respectively
\begin{eqnarray}
p^{\ast}_1 & = & \frac{1}{\sin{\varpi t}} \left ( q^2_{\ast} \cos{\varpi t } -
q^2_{\ast 0} \right )\, , \nonumber\\
p^{\ast}_2 & = & \frac{1}{\sin{\varpi t}} \left ( q^1_{\ast} \cos{\varpi t} -
q^1_{\ast 0} \right )\, , \nonumber\\
- p^{\ast}_{10} & = & \frac{1}{\sin{\varpi t}} \left ( q^2_{\ast 0}
\cos{\varpi t} - q^2_{\ast} \right ) \, , \nonumber\\
- p^{\ast}_{20} & = & \frac{1}{\sin{\varpi t}} \left ( q^1_{\ast 0}
\cos{\varpi t} - q^1_{\ast} \right ) \, ,
\end{eqnarray}
from which, together with (\ref{inv3}) and $Q^a=0$, $P_a=0$, $a=1,2$, the
solution (\ref{solution}) is obtained.

\section{Parametrizing the system}
If the Newtonian time $t$ is considered as configuration variable then one has
\begin{eqnarray}
S[x^\mu, t ] & = & \int^{\tau_2}_{\tau_1} d \tau \left [ \theta_{\mu} (x)
{\dot x}^{\mu} - H (x,t) \, {\dot t} \right ] \, ,
\end{eqnarray}
where the dot ``$\cdot$" stands for the time derivative with respect to the
unphysical parameter $\tau$. The next step is to define the momenta
$\pi_{x^{\mu}}$ canonically conjugate to $x^{\mu}$ and $\pi_t$ canonically
conjugate to $t$; respectively. By definition,
\begin{eqnarray}
\pi_{x^{\mu}} & := & \theta_{\mu} (x) \, , \quad \mu =1,2, \ldots , 2n\, , \\
\pi_t &:= & - H (x,t) \, ,
\end{eqnarray}
which lead to the primary constraints
\begin{eqnarray}
\chi_{\mu} & := & \pi_{x^{\mu}} - \theta_{\mu} (x) = 0 \, , \quad
\mu = 1,2, \ldots , 2n \, , \\
\gamma & := & \pi_t +  H (x,t) = 0 \, .
\end{eqnarray}
The next step is to compute the canonical Hamiltonian $H_c$
\begin{eqnarray}
H_c := \pi_{x^{\mu}} {\dot x}^{\mu} + p_t {\dot t } - L = \pi_{x^{\mu}} {\dot
x}^{\mu} + p_t {\dot t } - \left ( \theta_{\mu} (x) {\dot x}^{\mu} - H (x,t)
\, {\dot t} \right ) = 0 \, ,
\end{eqnarray}
in agreement with the fact that the system has been parametrized. Therefore,
\begin{eqnarray}
S [ x^{\mu} , t , \pi_{x^{\mu}} , \pi_t , \Lambda^{\mu} , \Lambda ] & = &
\int^{\tau_2}_{\tau_1} d \tau \left [ {\dot x}^{\mu} \pi_{x^{\mu}} + {\dot t}
\pi_t - \Lambda \gamma - \Lambda^{\mu} \chi_{\mu} \right ] \, .
\end{eqnarray}
The evolution with respect to $\tau$ of the primary constraint $\chi_{\mu}$ is
\begin{eqnarray}
{\dot \chi}_{\mu} & = & {\dot \pi}_{x^{\mu}} - \frac{\partial
\theta_{\mu}}{\partial x^{\nu}} {\dot x}^{\nu} \approx 0 \nonumber \\
& = & - \Lambda \frac{\partial H}{\partial x^{\mu}} + \omega_{\mu\nu} (x)
\Lambda^{\nu} \approx 0 \, ,
\end{eqnarray}
from which
\begin{eqnarray}\label{la}
\Lambda^{\mu} \approx \Lambda \omega^{\mu\nu} \frac{\partial H}{\partial
x^{\nu}} \, .
\end{eqnarray}
Similarly, the evolution with respect to $\tau$ of $\gamma$ is
\begin{eqnarray}
{\dot \gamma } & = & {\dot \pi}_t + {\dot H} = {\dot \pi}_t + \frac{\partial
H}{\partial x^{\nu}} {\dot x}^{\nu} + \frac{\partial H}{\partial t} {\dot t}
\nonumber\\
& = & - \Lambda \frac{\partial H}{\partial t} + \frac{\partial H}{\partial
x^{\nu}} \Lambda^{\nu} + \frac{\partial H}{\partial t} \Lambda \nonumber\\
& \approx & 0 \, ,
\end{eqnarray}
because of Eq. (\ref{la}). No more constraints arise. Therefore, the
constraints $\chi_{\mu}$ are second-class while $\gamma$ is related with the
first-class constraint
\begin{eqnarray}
G := \gamma + \chi_{\mu} \omega^{\mu\nu} (x) \frac{\partial H}{\partial
x^{\nu}} \, .
\end{eqnarray}
In fact,
\begin{eqnarray}
\{ G , \chi_{\mu} \} = \chi_{\alpha} \{ \omega^{\alpha\beta} \frac{\partial
H}{\partial x^{\beta}} , \chi_{\mu} \} \approx 0 \, ,
\end{eqnarray}
which has the usual structure between first-class and second-class constraints
\cite{henneaux}.

In this case, one has a dynamical system described by first-class and
second-class constraints. Due to the fact, one is dealing essentially with the
same physical situation, the Hamilton-Jacobi theory of the this case must lead
to the Hamilton-Jacobi equation given in Eq. (\ref{hjequation}). This is
indeed the case \cite{merino}.

\section{Conclusions}
In this paper we have made a proposal for the Hamilton-Jacobi theory for
unconstrained Hamiltonian systems described by symplectic structures and
Hamiltonians alternative to the usual ones. Our strategy consists in the
systematic application of Dirac's method to the action associated with such a
unconstrained Hamiltonian systems which leads to dynamical systems whose
extended phase space is endowed with a canonical symplectic structure and
second-class constraints. To handle the second-class constraints, we follow
essentially the procedure of Ref. \cite{rothe} to build the Hamilton-Jacobi
theory. It is important to emphasize that there exists another way,
alternative to the procedure of Ref. \cite{rothe}, of handling the
second-class constraints which consists in replacing the second-class
constraints by an equivalent set of first-class constraints enlarging the
phase space of the system following the Batalin-Tyutin procedure
\cite{batalin,amorim}. Once this has been achieved, the problem consists in
building the Hamilton-Jacobi theory for systems with first-class constraints,
which is well-known \cite{henneaux}.

Finally, it is also important to mention that the original problem of building
the Hamilton-Jacobi theory for unconstrained Hamiltonian systems described by
symplectic structures and Hamiltonians alternative to the usual ones can be
solved by means of Darboux's theorem \cite{Berndt}. In this framework, one
simply rewrites the original Hamiltonian system in terms of canonical
variables. Once this has been achieved, the Hamilton-Jacobi theory is the
usual one \cite{Berndt}. In this paper, however, we have avoided such a
procedure because the systematic application of Dirac's method seems to be the
natural procedure.

\section*{Acknowledgements}
The authors wish to thank G. F. Torres del Castillo and J. D. Vergara for very
fruitful discussions. This work was supported in part by the CONACyT grant
SEP-2003-C02-43939.


\end{document}